# MAS2HP: A Multi-Agent System to Predict Protein Structure in 2D HP Model

Hossein Parineh , Nasser Mozayani

**Abstract**

*Objectives* To develop a new solution for protein structure prediction (PSP) problem, by using "Agent Based Modeling"(ABM) as a computational method, in two dimensional hydrophobic-hydrophilic lattice model, based on previous works on this field.

*Results* A novel set of rules for Multi Agent System has been developed to predict 2D structure of protein based on its primary sequence. We broke the whole process of protein structure prediction into two steps: the first step, which was introduced in our previous article, is about biasing the linear sequence of protein to acquire a primary energy, and the next step, which is explained in this article, is about using ABM with a predefined set of rules, to find the best conformation in the least possible amount of steps and time. This method succeeded to find final 2D structure of protein for several benchmark sequences.

*Conclusions* Our suggested ABM model with new set of rules proved to be effective in finding the best conformations of benchmark sequences in remarkably shorter time compared with similar works.

*KeyWords*   Protein Structure Prediction . Agent Based Modelling . Multi Agent System . HP-model . 2D Lattice . Computational biology . NETLOGO

## Introduction

Proteins are very important molecules, which form the basis of many structures in nature. They are the building block and functional molecules in human body that play a key role in nearly every biological process. Almost all cellular processes on earth are governed or guided by proteins (Ghélis 2012). Proteins are made up of only 20 different types of amino acids, and the specifications and functions of each protein is determined not only by the type and quantity of its amino acids, but also by their sequence. The function of Protein is determined by its structure, and structure of a protein is determined by the sequence of its amino acids which are defined in the genetic code (Kessel et al. 2010).

In protein structure studies, the single most important research problem is to understand how protein sequences fold into their native 3D structures. This structure is determined by a complex set of interactions between both the amino acids themselves and the environment they are in (usually water molecules and other proteins) in a process known as folding. The problem of using sequence input to generate 3D structure output is referred to as the *ab-initio* protein structure prediction (PSP) problem and has been shown to be NP-Complete in both two-dimensional and three-dimensional square lattices (Berger et al. 1998). Accuracy of the predicted structure is evaluated by the measure of final energy of the structure. There is a huge gap between our capacities to extract protein sequences from DNA and to determine their 3-D structures. There are several experimental methods such as X-ray diffraction crystallography and nuclear magnetic resonance (NMR) but these methods are too slow and expensive. That's why there is an urgent need to develop a computational method to predict protein's structure based on its primary (linear) sequence, in a fast and reliable way.

In this article, we use *ab-initio* modeling wich is based on the *Anfinsen thermodynamic hypothesis* (Anfinsen 1972) that states "the *native* conformation adopted by a protein is the most stable one", i.e. the one with minimum free energy. In its

H.Parineh, N.Mozayani (✉)
School of Computer Engineering
Iran University of Science and Technology

Narmak,Tehran,Iran
Email: Parineh@alumni.iust.ac.ir , Mozayani@iust.ac.ir

native conformation, protein structure is thermodynamically stable, which shows that it complies with Gibbs second law of lowest free energy (second law of thermodynamic) (Anfinsen et al. 1961). We used 2D HP as the simplified model and multi-agent system as the computational method and a new approach for agent interactions.

## Materials and methods

Protein Structure

Protein structure is classified into four levels: primary, secondary, tertiary and quaternary. They refer to different stages of protein's formation. The primary structure is the sequence of amino acids that the protein is made up of. Secondary structure, called Alpha-Helix and Beta-sheet, are made by patterns of hydrogen bonds between the main chain peptide groups. Tertiary structure refers to the 3D structure of a single, double, or triple bonded protein molecule. Quaternary structure is the three dimensional structure of a multi subunit protein and how the subunits fit together. Main subject in PSP is finding an algorithm for the process of folding, i.e. reaching the tertiary sequence from primary sequence.

Protein Model

Due to restrictions in processing power of the current computer systems, all-atom folding simulations are unpractical. The protein structure prediction (PSP) problem could be modeled as an optimization problem in which the energy functions has to be minimized. As a result, several simplified models such as AB, HP, BLN and Tube Model have been proposed. Yet even simple abstractions are NP-complete (Berger et al. 1998). Dominant driving force for protein folding is the so-called hydrophobic force, consequently, regarding the fact that all 20 amino acids differ in their hydrophobicity, they could be classified to two groups: hydrophobic and polar (Li et al. 1997). The simplest model of protein is "HP lattice model" (Lau et al. 1989). Later 2D HP Model was introduced (Lau et al. 1990) and has been intensively studied (Li et al. 1996 ; Shmygelska 2003; Liu et al. 2014). It models a protein as a linear chain of amino acid residues. Each amino acid can be either of two types: H (hydrophobic, i.e., nonpolar) or P (hydrophilic, i.e., polar). For simplicity, we denote H by "1" (black) and P by "0" (white).

Energy Function

To assess algorithm effectiveness in different methods and models, final energy of the predicted structure is compared. In this work, we use 2D HP lattice model. The energy for a sequence folded into a structure in 2D/3D HP model is given by the short range contact interactions:

$$E(S) = \sum_{i<j} e_{v_i v_j} \Delta(r_i - r_j)$$

where $\Delta(r_i - r_j) = 1$ if $r_i$ and $r_j$ are adjoining lattice sites but i and j are not adjacent in position along the sequence, and $\Delta(r_i - r_j) = 0$ otherwise. Depending on the types of monomers in contact, the interaction energy $e_{v_i v_j}$ will be $e_{HH}$, $e_{HP}$, or $e_{PP}$, corresponding to H-H, H-P, or P-P contacts, respectively (Fig.1-b).

$$e_{v_i v_j} = \begin{cases} -1 & for\ e_{HH} \\ 0 & otherwise \end{cases}$$

NETLOGO

Netlogo is a programmable modeling environment designed for simulating complex natural and social phenomena. Modelers can give instructions to hundreds or thousands of "agents" all operating independently. This makes it possible to explore the connection between the micro-level behavior of individuals and the macro-level patterns that emerge from their interaction.

Environment

By evaluating the problem definitions, the environment is "accessible", "deterministic", "episodic", "static" and "discrete"(Wooldridge, 2009).

MAS2HP structure

During last three decades different classes of methods for ab-initio modeling such as: Ant Colony Optimization (ACO) (Shmygelska 2003), Evolutionary algorithms (EA) /Genetic Algorithm (GA) (Krasnogor et al. 1999), tabu search algorithm (Lin et al. 2014), Multi Agent Systems (MAS) (Lipinski-Paes et al. 2016) and etc. have been proposed. These methods were implemented on either 2D or 3D HP model. Agents in MAS2HP are defined in three different levels. Higher level agents are responsible to supervise the structure and other agents, while lower level agents are mainly responsible to search through confirmation space. Environment agent, as human-machine-interface agent, draws diagrams and interacts with operator. Each level will be described in a hierarchical bottom-up order. The way agents interact is shown in figure 1-c. Level 1 or searching agents: Each amino acid is represented by an agent $A_i$ and owns $\{Type_i, PosX_i, PosY_i\}$ attributes. These agents are mainly responsible for searching through conformation space based on information that is provided by other agents about their type (H or P) and position (x and y). They also share their movement information with other agents. Main information for selecting a destination to move by an agent are "H_center" and "CP_list" which is calculated by supervisor agent. Level 2- Supervisor agent: In this level there is a supervisor agent, who is mainly responsible for monitoring the structure built by level-1 agents and as soon as the structure encounters a conflict, it will turn everything to the last best conformation predicted before. It also computes energy of the structure and temperature of the environment. This agent makes two important calculations based on a table called "HP_List" updated by level-1 agents. First level agents report their position to this agent and it calculates virtual center of all H type agents. It is used to determine center of the structure. Next important calculation is about updating a table of possible places for H-H connection (which are available and won't lead to dead-end in sequence). Later on, agents on level-1 would select between possible places for them to move on. Level 3- Environment agent: This agent is mainly responsible to draw different diagrams about temperature of the environment and energy of the structure. It is also used to communicate with operator and other agents.

RESTRICTIONS

Regardless of selected model for PSP, two conditions must be met. Let suppose the primary sequence of a protein $P$ is a sequence $S$ of n amino acids (Hydrophobic/Hydrophilic):
$S = S_1, S_2, \ldots, S_n$ where $S_i \in \{H, P\}, \forall\, 1 \leq i \leq n$

First rule (Self avoidance): The first rule explains that two amino acids (agents) must not coincide on the same position on the lattice plane:
$$\forall S_i \in S, \forall S_j \in S, with\ i \neq j \rightarrow$$
$$(posX_i, posY_i) \neq (posX_j, posY_j)$$

Second rule (Connected neighbors): The distance between two consecutive amino acids (agents) on the sequence must not exceed 1, i.e. they should be neighbors on the lattice.
$$\forall\, 1 \leq i, j \leq n, with\ |i - j| = 1 \rightarrow$$
$$|X_i - X_j| + |Y_i - Y_j| = 1$$

Movement Direction
In HP-Model, considering its lattice structure, agents could move in four directions i.e. north, south, west and east. In addition to these directions we also use diagonal movement which was first introduced in (Lin et al. 2014). In our work, the first agent may only move in diagonal directions i.e. NE, NW, SE and SW, while other follower agents could move in all directions i.e. N, NE, NW, W, E, SE, SW and S (Figure 1-d).

Movement Strategy
Agents in a system may work simultaneously or sequentially. In this work agents move sequentially. First an agent moves to its best "accessible" and "possible" place. After that, regarding the rule no.2 in previous section, other agents tend to keep their legal distance with the moved agent and they will move toward it. These new movements may lead to other agents to loose their legal distance with the newly moved agents. As a result these chain movements will cease only when inter-agent distances, all over the structure are equal to 1(figure 1-a).

Control Folding Process
In our previous article (Mozayani and Parineh 2015), we presented a method called FBA (Fast Bias Algorithm) which was able to receive a linear sequence of amino acids (primary structure of a protein) and bias it to an elementary version of 2D structure. As a result, the linear sequence with zero energy would turn to a 2D structure (not the best conformation) with a better energy level in a very short time. Now, with MAS2HP we want to find the best possible structure of the protein which means the best energy of the conformation. In every step of running MAS2HP, as agents move, the structure would change from $S_1$ to $S_2$. This change in structure will alter the amount of energy from $E_1$ to $E_2$. It is important to know that facing worse energies is not always bad. Especially in the commence of simulation it would help to evade from local minima and search in wider range of conformation space. As simulation continues, the range of conformation space will decrease, and the final answer will be found in a more limited conformation space. In order to achieve it, we use SA (Simulated Annealing) technic. SA is a simple and effective meta-heuristic algorithm. Here, at the start of simulation, first energy of the structure which is gained by implementing FBA (Parineh and Mozayani 2015) on a linear (primary) sequence, will be considered as

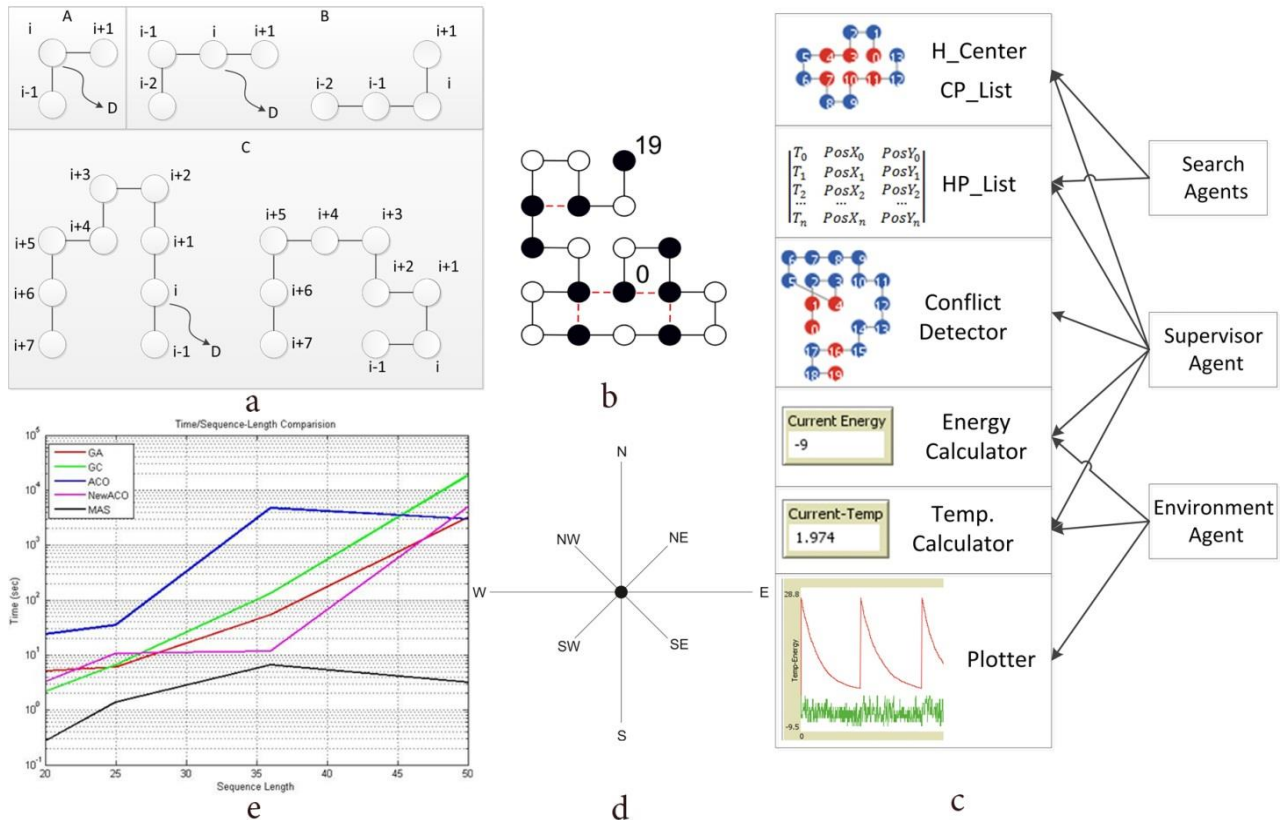

**Fig 1**: a) Movement Strategy of leader and follower agents. b) displaying H-H connection which is used to calculate energy of the structure. c) Structure of MAS2HP and access level of agents. d) Possible moving direction for agents. e) Comparison of elapsed time for each algorithm to reach the final structure. While the length of sequence increases the time is exponential for all algorithms except MAS2HP.

the current energy and current structure be will considered as the last best conformation. Then agent movement starts. After a new move of an agent, the energy of the structure will change in a better or worse state. Provided that $E_1 < E_2$, the last move of the agent will be accepted, current structure will be set as last best conformation and current energy would be set as last best energy. But if $E_1 > E_2$, there is still a possibility to accept new move. The probability of accepting new move is $\exp(-\Delta E/T)$, where $\Delta E$ is the difference between current energy and last best energy of the system and $T$ is temperature of the system. At the start of simulation, overall temperature of the system is high, and as simulation continues, it is decreased. As a result, in final steps, there is a small chance to get worse answer thus the algorithm will converge to an acceptable answer (low-energy conformation). Temperature will change with a constant rate, $\alpha$ which is defined by user:

$T_{new} = \alpha . T_{old}$ , $\alpha = 0.98$

Simulation will stop when temperature of the environment decreases under a certain value that is defined by the user. At the end of simulation, last best conformation and last best energy will be selected and displayed as the final answer.

**Results and discussion**
Implementing in NETLOGO
In this section MAS2HP will be introduced. Figure 2 displays the software interface. MAS2HP is implemented in NETLOGO. The first step in using MAS2HP is giving it the linear sequence of protein in hydrophobic-hydrophilic format. Then it should be initialized with setup button. "Bias" button is used to implement FBA algorithm (Mozayani and Parineh 2015), and "MAS" button is used to implement MAS2HP method on the sequence. There are several buttons, slide buttons, switches and displayers that most important of, them will be introduced briefly. In the startof simulation decrease rate of the temperature, knowm as "$\alpha$", and the temperature simulation termination is set by user. In display section, current energy and temperature of the system in addition to elapsed time to reach to reach current structure is displayed. Furthermore, a plot that displays states of energy and temperature during simulation is presented.

**Table 1** Comparing the results of implementing different algorithm on first group of benchmark protein sequences based on final energy of the predicted strcuture

| No | Length | $E_{MAS}$ | $E_{GA}{}^{*a}$ | $E_{MMC}{}^{*b}$ | $E_{CG}{}^{*C}$ | $E_{ACO}{}^{*D}$ | $E_{PFGA}{}^{*e}$ | Sequence |
|---|---|---|---|---|---|---|---|---|
| 1 | 20 | -9 | -9 | -9 | -9 | -9 | -9 | HPHPPHHPHPPHPHHPPHPH |
| 2 | 25 | -8 | -8 | -8 | -8 | -8 | -8 | PPHPPHHPPPPHHPPPPHHPPPPHH |
| 3 | 36 | -14 | -14 | -13 | -14 | -14 | -14 | PPPHHPPHHPPPPPHHHHHHHPPHHPPPPHHPPHPP |
| 4 | 50 | -19 | -21 | -21 | -21 | -21 | -21 | HHPHPHPHHHHHPHPPPHPPPHPPPPHPPHPPPHPHHHHPHPHPHPHH |

*a and *b respectively by using "Genetic Algorithm" and "Metropolis Monte Carlo" method (Unger et al. 1993). *c: Using "Chain Growth" method (Betler et al. 1996). *d: By using "Ant Colony Optimization" method (Shmygelska 2003). *e: By using "Protein Folding Genetic Algorithm" method (Bui et al. 2005)

**Table 2** Comparing the results of implementing different algorithm on second group of benchmark protein sequences based on final energy of the predicted strcuture

| No | Length | $E_{Mas}$ | $T_{Mas}$ (s) | MAS Runs | $E_{GA}{}^{*a}$ | Sequence |
|---|---|---|---|---|---|---|
| 1 | 18 | -9 | 2.271 | 3 | -9 | PHPPHPHHHPHHPHHHH |
| 2 | 18 | -8 | 0.927 | 2 | -8 | HPHPHHHPPPHHHHPPHH |
| 3 | 18 | -4 | 2.607 | 6 | -4 | HHPPPPPHHPPPHPPPHP |
| 4 | 20 | -10 | 1.445 | 1 | -10 | HHHPPHPHPHPPHPHPHPPH |

*a: By using "Genetic Algorithm" method (Krasnogor et al. 1999)

**Table 3** Comparing the elapsed time to reach final energy of the structure for different methods

| | | | Time (seconds) | | | |
|---|---|---|---|---|---|---|
| No | Length | $T_{MAS}$ | $T_{GA}{}^{*a}$ | $T_{CG}{}^{*b}$ | $T_{ACO}{}^{*c}$ | $T_{NEWACO}{}^{*d}$ |
| 1 | 20 | 0.272 | 5.16 | 2.16 | 23.90 | 3.33 |
| 2 | 25 | 1.384 | 6 | 6.6 | 35.32 | 10.62 |
| 3 | 36 | 6.655 | 54.6 | 132 | 4746.12 | 11.81 |
| 4 | 50 | 3.181 | 3180 | 18600 | 3000.28 | 4952.92 |

*a,*b: By using "Genetic Algorithm" and "Chain Growth" method (Beutler et al. 1996)

Result of Simulation

To evaluate algorithm efficiency, it was tested by eight benchmark sequences with different lengths and was compared with other algorithms in terms of energy and elapsed time to reach final conformation. These benchmark sequences are displayed in table 1 and table 2. In the first step, energy of final structure ($E_{MAS}$), acquired by implementing MAS2HP on protein's primary sequences from table 1 is compared with other articles. MAS2HP was completely successful in finding the best conformation for sequences 1 to 3, and for the sequence number 4 with 50 amino acids, the best energy of the structure was -19 (table 3) . Now, sequences of table 2 will be tested by MAS2HP and the results will be compared with "Genetic Algorithm" from (Krasnogor et al. 1999) and displayed in table 2. However number of tries and spent time to reach each final conformation is significantly small, MAS2HP reached the same results of other algorithms. By comparing elapsed time between these algorithms to reach final structure, it is obvious that our algorithm is doing much better in terms of reaching final structure in a significantly shorter time, but more important result is that MAS2HP had changed the exponential time used for PSP process to a linear time, as displayed in figure 2. To reach these results, MAS2HP was implemented 25 times on each sequence, in a core-i3 2.5GHz CPU with 3G RAM, and the best answer was selected.

**Conclusion and future work**

New method of using ABM in PSP, by implementing a new structure for inter-agent and agent-environment communication, and agent movement direction and strategy, MAS2HP proved to be effective in finding the best conformation of the benchmark sequences provided in table 1 and 2, except one example. It reached the best conformations in a considerably shorter time compared to other algorithms. The reliability of using MAS2HP to find the best conformation for HP model of protein 2D lattice model, was tested for over 10 tries, each included 25 run for each sequence, and the best result between each 25 run was selected for each try. However the best energy/time was selected from each 25 runs, in every 10 tries different structures with the best energy and significantly shorter time, regarding other algorithms, was achieved which shows its performance is reliable.

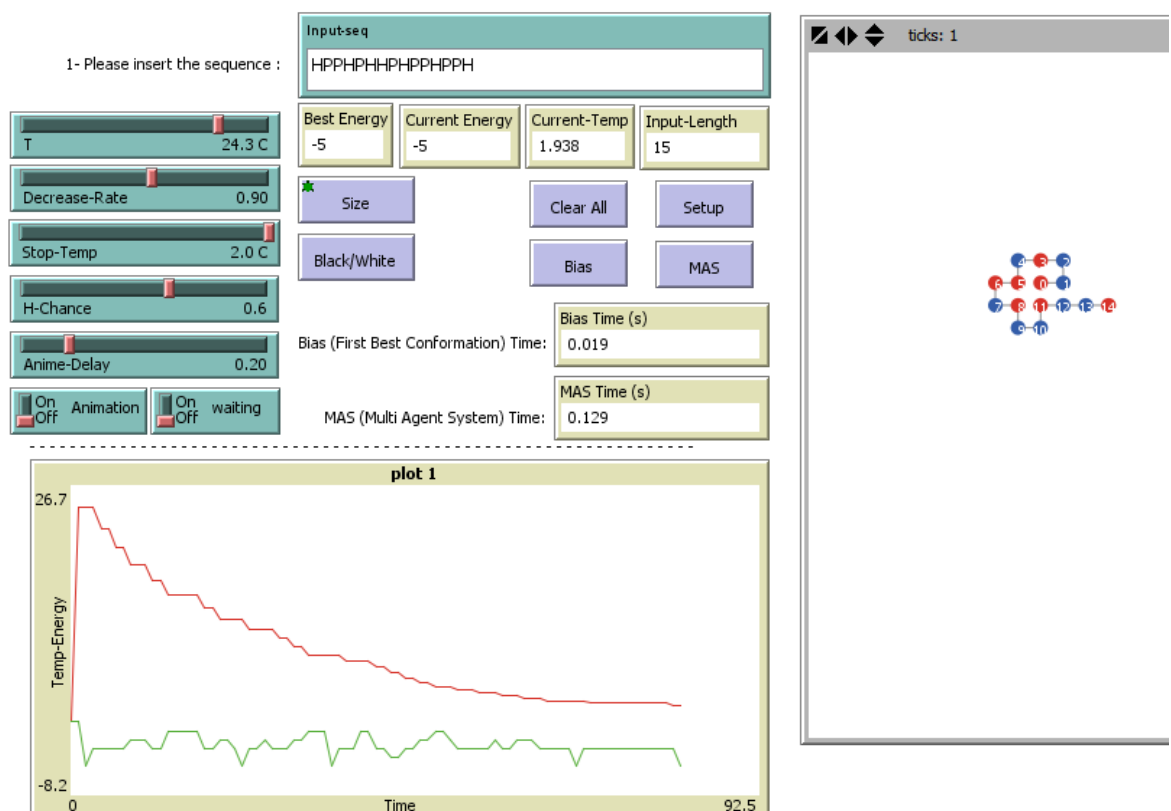

**Fig. 2**: MAS2HP interface. In MAS2HP user inserts linear sequence of amino acids in the software. Then by selecting "setup" button it is shown in display. Next step is biasing the linear sequence which is done by "Bias" button. Then MAS2HP algoirthm is implemented by "MAS" button. An overview of temprature (red) and energy (green) of the system is plotted. Elapsed times for each step are displayed too. Other settings are possible by the buttons in software.

Regarding the fact that this method doesn't use memory to enhance its performance by using previous prediction processes experiments, it is suggested to add memory in next versions. It is also believed that using "Re-inforcement Learning" could improve its performance. Some modifications in algorithm and structure of movement would enhance the result of this method's performance.